\documentclass[aps,twocolumn,pra,superscriptaddress,amsmath,showpacs,tightenlines]{revtex4}
\usepackage{epsfig,graphicx,times}
\usepackage{amstext}
\usepackage{amsmath}            %serve per le subequazioni
\usepackage{amssymb}            %serve per il simbolo "marchio registrato", \circledR
\usepackage{graphicx}           %serve per le figure eps, ps etcyy
\usepackage{mathrsfs}
\usepackage{bm}
\usepackage[tight]{subfigure}
\usepackage[colorlinks,linkcolor=blue,anchorcolor=blue,citecolor=blue,urlcolor=black]{hyperref}

\begin{document}

\title{Optomechanical second-order sidebands and group delays in a Kerr resonator}

\author{Ya-Feng Jiao}
\affiliation{Department of Physics, Henan Normal University, Xinxiang 453007, China}
\affiliation{Key Laboratory of Low-Dimensional Quantum Structures and Quantum Control of Ministry of Education, Department of Physics and Synergetic Innovation Center for Quantum Effects and Applications, Hunan Normal University, Changsha 410081, China}

\author{Tian-Xiang Lu}
\affiliation{Key Laboratory of Low-Dimensional Quantum Structures and Quantum Control of Ministry of Education, Department of Physics and Synergetic Innovation Center for Quantum Effects and Applications, Hunan Normal University, Changsha 410081, China}

\author{Hui Jing}\email{jinghui73@gmail.com}
\affiliation{Key Laboratory of Low-Dimensional Quantum Structures and Quantum Control of Ministry of Education, Department of Physics and Synergetic Innovation Center for Quantum Effects and Applications, Hunan Normal University, Changsha 410081, China}

\date{\today}

\begin{abstract}
We theoretically study high-order optomechanically-induced transparency (OMIT) process in a nonlinear Kerr resonator. A frequency shift induced by the Kerr effect, is identified for the optical cavity mode, which results in asymmetric OMIT windows of the signal and its higher-order sidebands. We also find that both the sideband amplitude and its associated group delay sensitively depend on the strength of the Kerr nonlinearity. This indicates the possibility to enhance or steer the performance of OMIT devices with various nonlinear optical cavities.
\end{abstract}

\maketitle

Optical microcavities, with ultrahigh quality factor and small mode volume, provide giant enhancement of coherent light-matter interactions, which is important for many applications~\cite{Vahala424}, such as microlasers~\cite{Kippenberg740}, optical routing~\cite{Dayan319} or sensing~\cite{Zhu446}, and optomechanics (OM)~\cite{Aspelmeyer861,Aspelmeyer657,Kippenberg321,Marquardt240}. A recent advance closely related to the present study is optomechanically induced transparency (OMIT)~\cite{Agarwal810,Weis330,Safavi472}. As a solid-state analogy to electromagnetically induced transparency (EIT) originally observed in atomic gases~\cite{Fleischhauer776}, the fundamental OMIT features the two-channel destructive interference of the absorptions of the probe photons (by the cavity itself or the mechanical mode). Beyond this picture, high-order OMIT effects also emerge due to the intrinsic nonlinear OM interactions~\cite{Fan658,Lemonde111,Kronwald111}, such as photon-phonon polariton pairs~\cite{Liu111} and sideband generations~\cite{Xiong860}. OMIT not only provides an alternative approach for achieving quantum memories~\cite{Fiore107,Hill311,Dong338,Zhou917}, but also opens up the way to explore a variety of new effects, such as nonreciprocal OMIT~\cite{Shen106}, phase-tuned OMIT~\cite{Jia910}, active OMIT~\cite{Jing596}, two-color OMIT~\cite{Wang900} and OMIT with Bogoliubov phonons~\cite{Dong142}. In contrast, high-order OMIT sidebands are generally much weaker than the probe signal and thus hard to be detected or utilized. Hence the sideband enhancement becomes important for its potential applications in e.g., precise sensing of charges~\cite{Xiong423,Kong950} or weak forces~\cite{Borkje111}, single-particle detection~\cite{Li116}, magnetometer~\cite{Liu712}, and high-order squeezed frequency combs~\cite{Liu439}.

Very recently, nonlinear optical effects in materials such as parametric amplifications and optical Kerr effect, which can be strongly enhanced by the photons circulating in microcavities, have attracted intense studies~\cite{Boyd2008,Zhou421,Cao118,Mitchell425}. In particular, based on the enhanced Kerr effects~\cite{Cao118,Mitchell425,Lin982}, a number of important applications have been realized, including chiral-symmetry breaking~\cite{Cao118}, self-tuning resonator~\cite{Mitchell425}, all-optical switching~\cite{Yoshiki222}, solitons~\cite{Suh354,Guo139,Obrzud116}, frequency combs~\cite{Brasch351,Del450,Kippenberg332}, and nonlinear OM control~\cite{Mikkelsen960}. In a recent experiment with a high-Q silica microsphere, asymmetric Fano-like OMIT spectrum was observed due to the optical Kerr effect~\cite{Shen411}, which can be further tuned or compensated by varying the pump power and the optical frequency.

\begin{figure*}
	\centering
	\subfigure{
		\label{Fig1.sub.1}
		\includegraphics[width=0.32\textwidth]{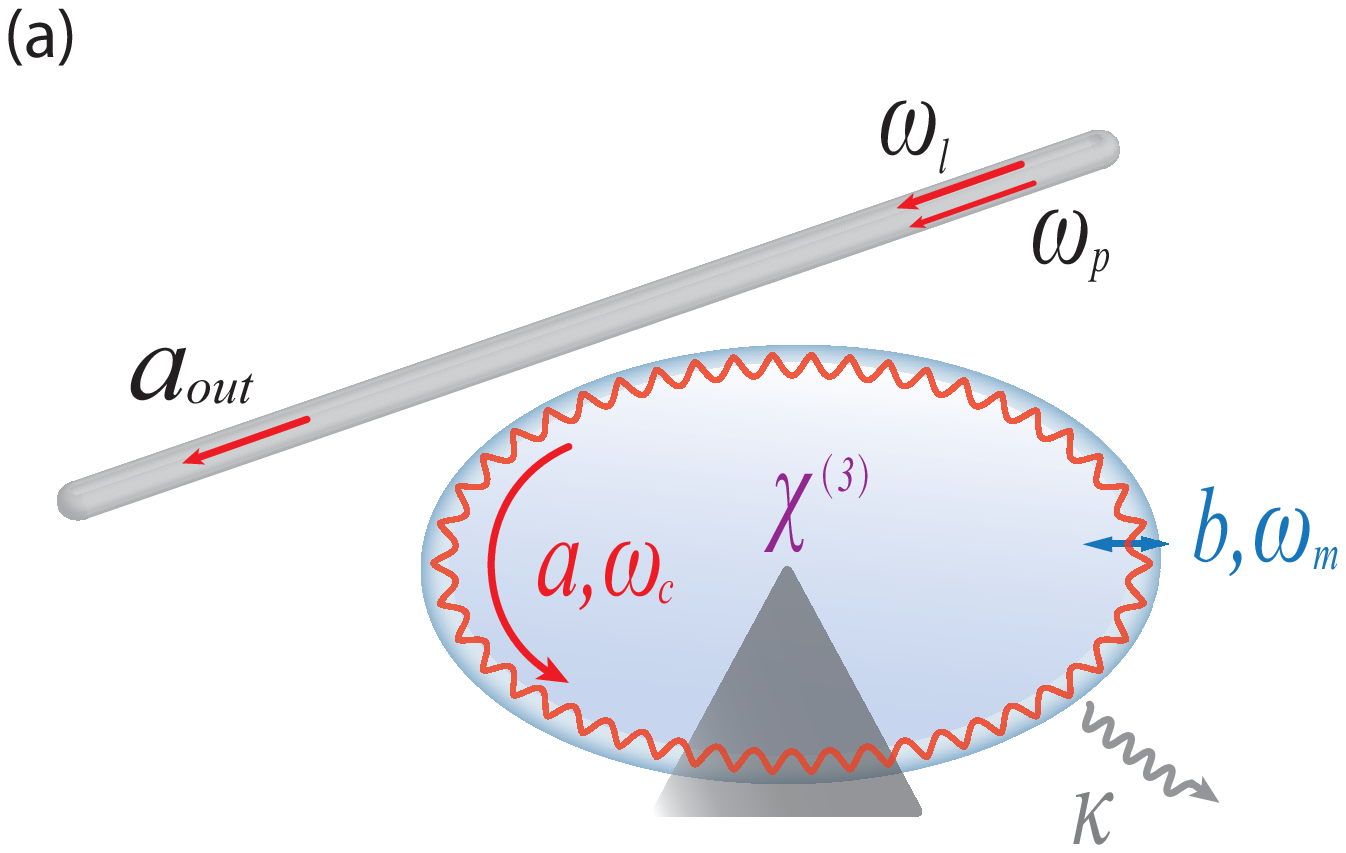}}
	\subfigure{
		\label{Fig1.sub.2}
		\includegraphics[width=0.32\textwidth]{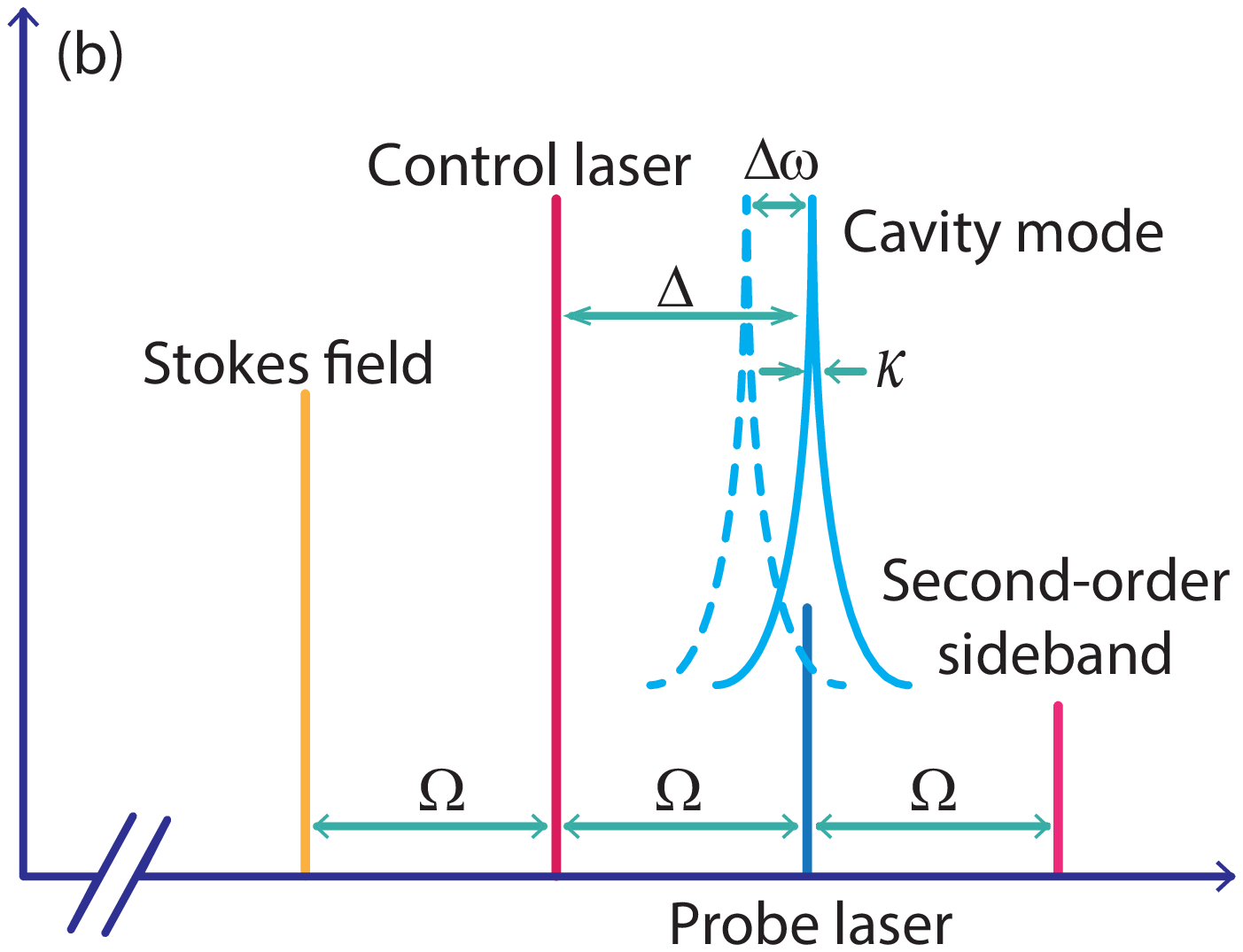}}
	\subfigure{
		\label{Fig1.sub.3}
		\includegraphics[width=0.32\textwidth]{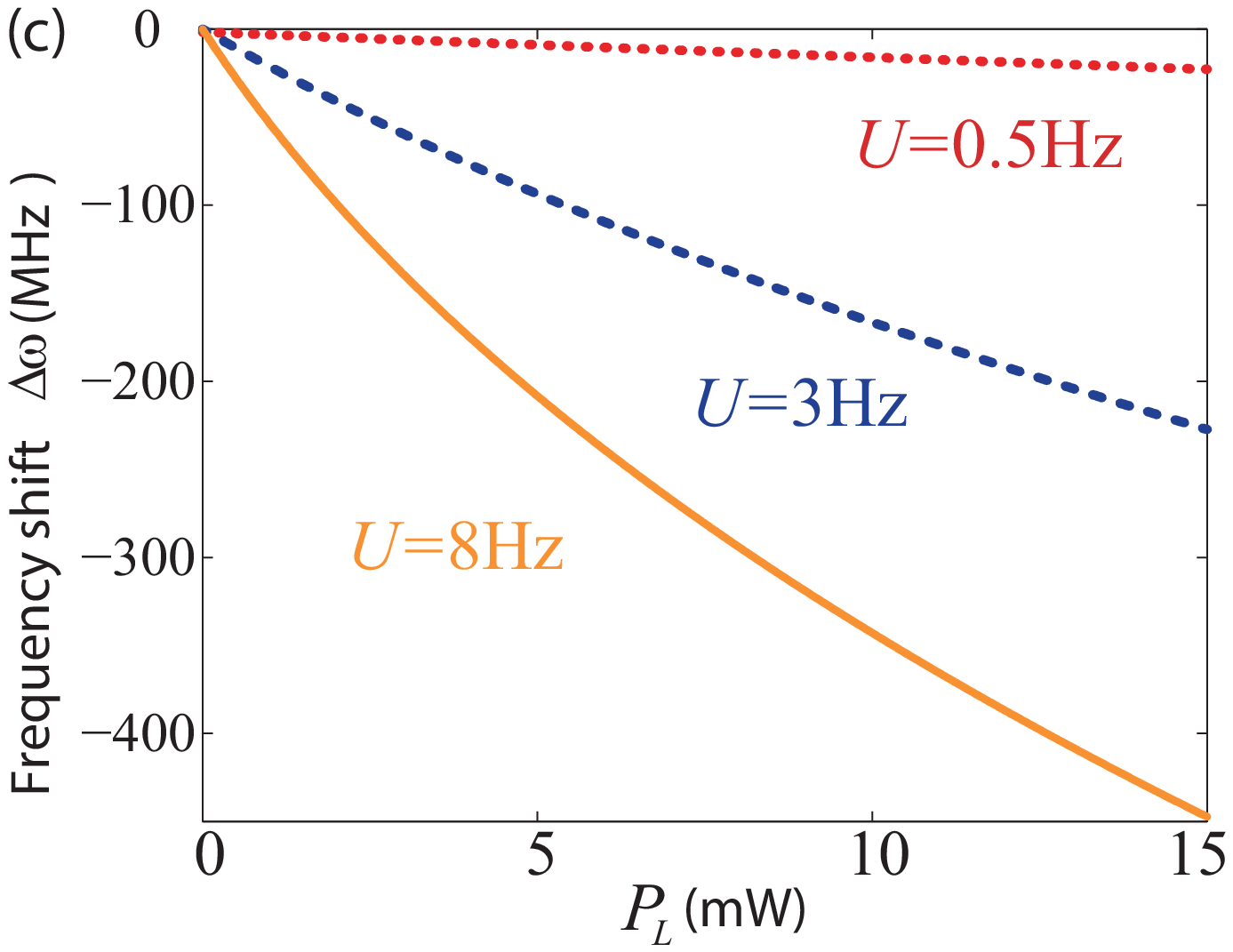}}	
	\caption{(Color online) (a) Schematic diagram of an OM system made of Kerr-type nonlinear material. The system is driven by a strong control field of frequency $\omega_{l}$ and a weak probe field of frequency $\omega_{\textrm{p}}$. (b) Frequency spectrum of the hybrid system. The dashed line shows that the cavity mode, which has a linewidth of $ \kappa $, is red-detuned by $ \Delta\omega $ due to the Kerr effect. (c) The frequency shift $\Delta\omega$ varies with the pump power $ \textrm{P}_\textrm{L} $ and Kerr coefficient $ U $.}
	\label{Fig.1}
\end{figure*}

In this paper, based on the OMIT experiment in a Kerr cavity~\cite{Shen411}, we proceed to study high-order OMIT and its associated group delay in such a nonlinear cavity. We find that in the presence of optical Kerr effect, compared to that in a linear resonator, the amplitude of second-order sideband can be significantly enhanced. This sideband amplitude also can be further tuned by the external light, since the Kerr-induced shift can be either compensated or amplified by varying the pump frequency. Moreover, the delay time of second-order sideband sensitively depends on the Kerr nonlinearity. At high pump power, the sideband can be tuned from fast light to slow light, which is potentially useful for optical storage or switch. The enhanced nonlinear OMIT in a Kerr resonator, as revealed here, opens up a promising new way to study other important OM effects, e.g., motion cooling~\cite{Huang790} or squeezing~\cite{Wollman349}, light-sound entanglement~\cite{Vitali980}, and photon blockade~\cite{Nunnenkamp107,Rabl107,Ferretti850}.

As shown in Fig.\,\ref{Fig1.sub.1}, we consider the nonlinear OMIT in a Kerr resonator. A pump laser of frequency $ \omega_{l} $ and a probe laser of frequency $ \omega_{p} $ are applied to the system via the evanescent coupling of the optical fiber and the resonator, the field amplitudes are given by 
\begin{align}
\nonumber
\varepsilon_{l}=\sqrt{\dfrac{\kappa P_{L}}{\hbar\omega_{l}}},~~~~~\varepsilon_{p}=\sqrt{\dfrac{\kappa P_{s}}{\hbar\omega_{p}}},
\end{align}
where $ P_{L} $ and $ P_{s} $ are the pump and probe powers, respectively. In the rotating frame at the pump frequency $ \omega_{l} $, the Hamiltonian of this OM system is given by (for $ \hbar=1 $):
\begin{align}
\nonumber
&H=H_{0}+H_{\textrm{int}}+H_{\textrm{dr}}, \\ \nonumber
&H_{0}=-\Delta_{c}\hat{a}^\dag \hat{a} + \dfrac{\hat{p}^2}{2m} +\frac{1}{2}m\omega_m^2\hat{x}^2,\\ \nonumber
&H_{\textrm{int}}= g\hat{a}^\dag \hat{a}\hat{x}- U \hat{a}^\dag \hat{a}^\dag \hat{a}\hat{a},\\ 
&H_{\textrm{dr}}=i\big(\varepsilon_{l}\hat{a}^\dag + \varepsilon_{p}\hat{a}^\dag e^{-i\Omega t} - H.c.\big), \label{eq1}
\end{align}
where $\hat{a}$ is the annihilation operator of the cavity field (with resonance frequency $\omega_{c}$), $\hat{p}$ and $\hat{x}$ describe the momentum and position of the mechanical mode with effective mass $m$ and frequency $\omega_{m}$. The nonlinear OM and Kerr coupling coefficients are denoted by $g$ and $U$. The pump-cavity and probe-pump detuning are, respectively, denoted by
\begin{align}
\nonumber
\Delta_{c}=\omega_{l}-\omega_{c},~~~~~\Omega=\omega_{p}-\omega_{l}.
\end{align}
When the pump and probe lasers are incident upon the cavity, high-order sidebands with frequencies $ \omega_{l}\,\pm\,n\Omega $ emerge due to nonlinear OM interactions, where $n$ is an integer representing the order of the sideband~\cite{Xiong860}. The frequency spectrum is shown in Fig.\,\ref{Fig1.sub.2}. The first-order sidebands with frequencies $ \omega_{l}\,\pm\,\Omega $ are referred to as the Stokes and anti-Stokes fields, respectively. The sidebands with frequencies $ \omega_{l}\pm2\Omega $ are the second-order upper or lower sidebands.

The nonlinear Kerr coefficient is usually given by~\cite{Brasch351}:
\begin{align}
U=\dfrac{\hbar \omega_{c}^2 c n_{2}}{n_{0}^2\textrm{V}_{\textrm{eff}}},\label{eq2}
\end{align}
with
\begin{align}
\textrm{V}_{\textrm{eff}}=\int_{V}\varepsilon(r)|\Phi(r)|^{2}dV,\label{eq3}
\end{align}
where c is the speed of light in vacuum, $ n_0 $ and $ n_2 $ are the linear and nonlinear refractive index of the material with typical values $ 2\leq n_{0}\leq4 $ and $ 10^{-14}\leq n_{2}\leq10^{-17} \textrm{cm}^{2}/\textrm{W} $~\cite{Boyd2008}. $\textrm{V}_{\textrm{eff}}$ is the effective mode volume, describing the peak electric field strength within the cavity. $ \varepsilon(r) $ and $ \Phi(r) $ are the dielectric constant and the electric field strength, respectively. For a high Q factor ($ 10^{6}\sim10^{8} $), $\textrm{V}_{\textrm{eff}}$ is typically between $10^{2}\sim10^{4}\,\mu\textrm{m}^{3}$~\cite{Vahala424}. 

For a near infrared wavelength (i.e. $ \lambda=780\,\textrm{nm} $), calculating Eq.\,(\ref{eq2}) with experimentally accessible parameters, the Kerr coupling coefficient $U$ is estimated on the order between $ 0.1\sim10$ Hz in a silica microsphere. Note that there is a red-shift of the cavity mode due to the influence of the Kerr effect~\cite{Shen411}, i.e., $ \omega_{c}\rightarrow \omega_{c}^{'}=\omega_{c}+\Delta\omega $ and
\begin{align}
\Delta\omega\equiv-2U|\bar{a}|^{2}=\dfrac{2Um\omega_{m}^{2}\bar{x}}{g},\label{eq4}
\end{align}
with
\begin{align}
\bar{a}=\dfrac{\varepsilon_l}{-i(\Delta_c-\Delta\omega-g\bar{x})+\kappa},\label{eq5}
\end{align}
where $\bar{a}$ and $\bar{x}$ are the steady-state values of the optical and mechanical modes, respectively. The frequency shift $ \Delta\omega $, being related to both the strength of the Kerr effect and the pump power, is shown in Fig.\,\ref{Fig1.sub.3}. For $U=0.5\,\textrm{Hz}$, $ \Delta\omega $ is well agreed with the observed result in Ref.\cite{Shen411}. For a fixed pump power $ P_{L} $, stronger Kerr nonlinearity enhances the values of both $ \Delta\omega $ (the optical-resonance shift) and $ |\bar{a}|^{2} $ (the intra-cavity photon number), which in turn can significantly modify the nonlinear OMIT process.

In this paper, we focus on the mean response of the system to the probe field, so the quantum noise can be ignored. In this case, we use semiclassical Heisenberg-Langevin equations to describe the evolution of this hybrid system. By introducing dissipation terms, the equations of motion (EOM) can be written as:
\begin{align}
\nonumber
&\dot{a}=(i\Delta_c-\kappa)a-igax+2iUa^\dag aa+\varepsilon_l+\varepsilon_pe^{-i\Omega t},\\
&m\ddot{x}+m\gamma_m\dot{x}+m\omega_m^2x=-ga^\dag a,\label{eq6}
\end{align}
where $\kappa$ and $\gamma_m$ are the decay rates of the optical and mechanical mode, respectively.

\begin{figure*}[htbp]
	\centering
	\subfigure{
		\label{Fig2.sub.1}
		\includegraphics[width=0.32\textwidth]{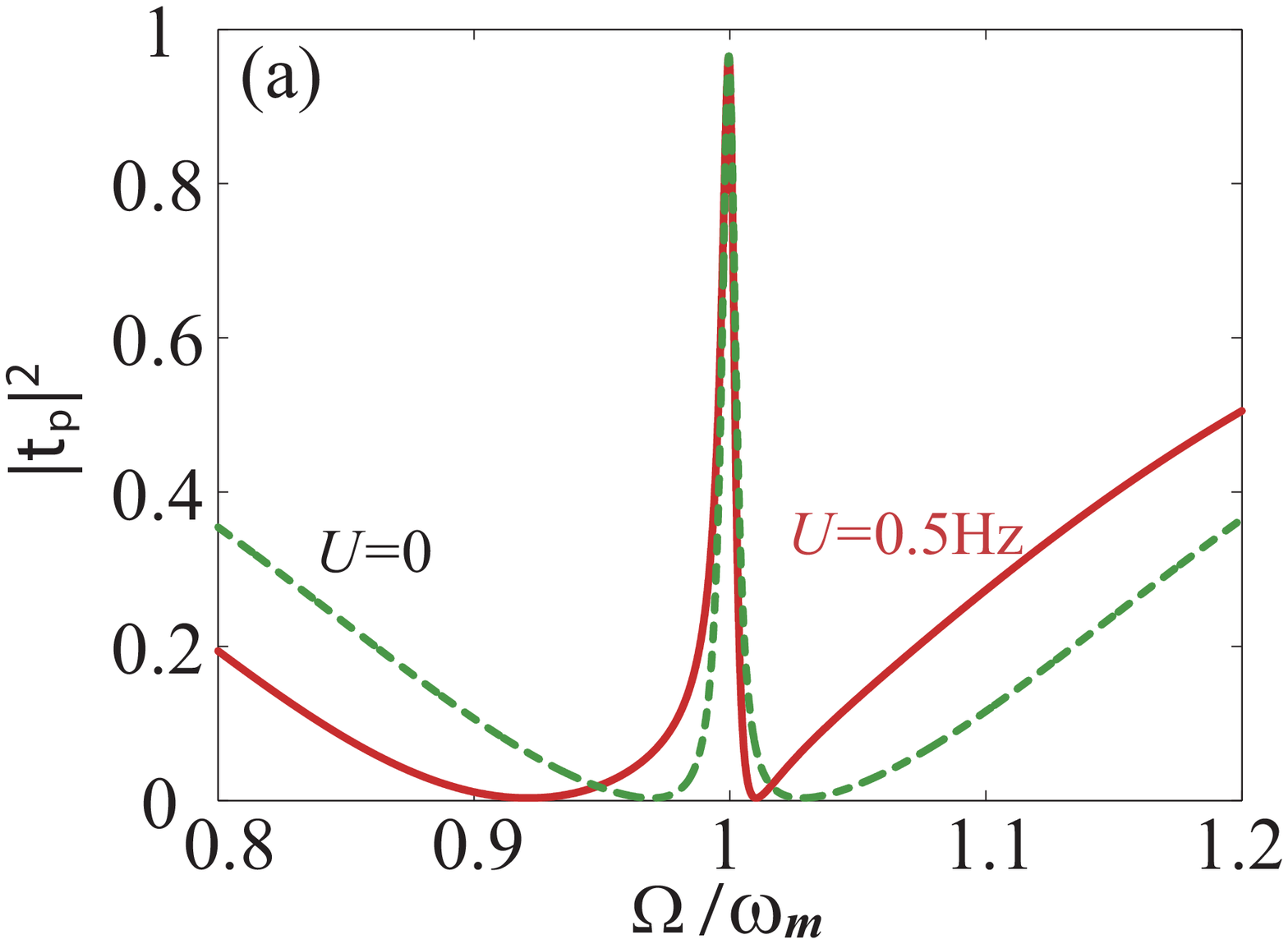}}
	\subfigure{
		\label{Fig2.sub.2}
		\includegraphics[width=0.32\textwidth]{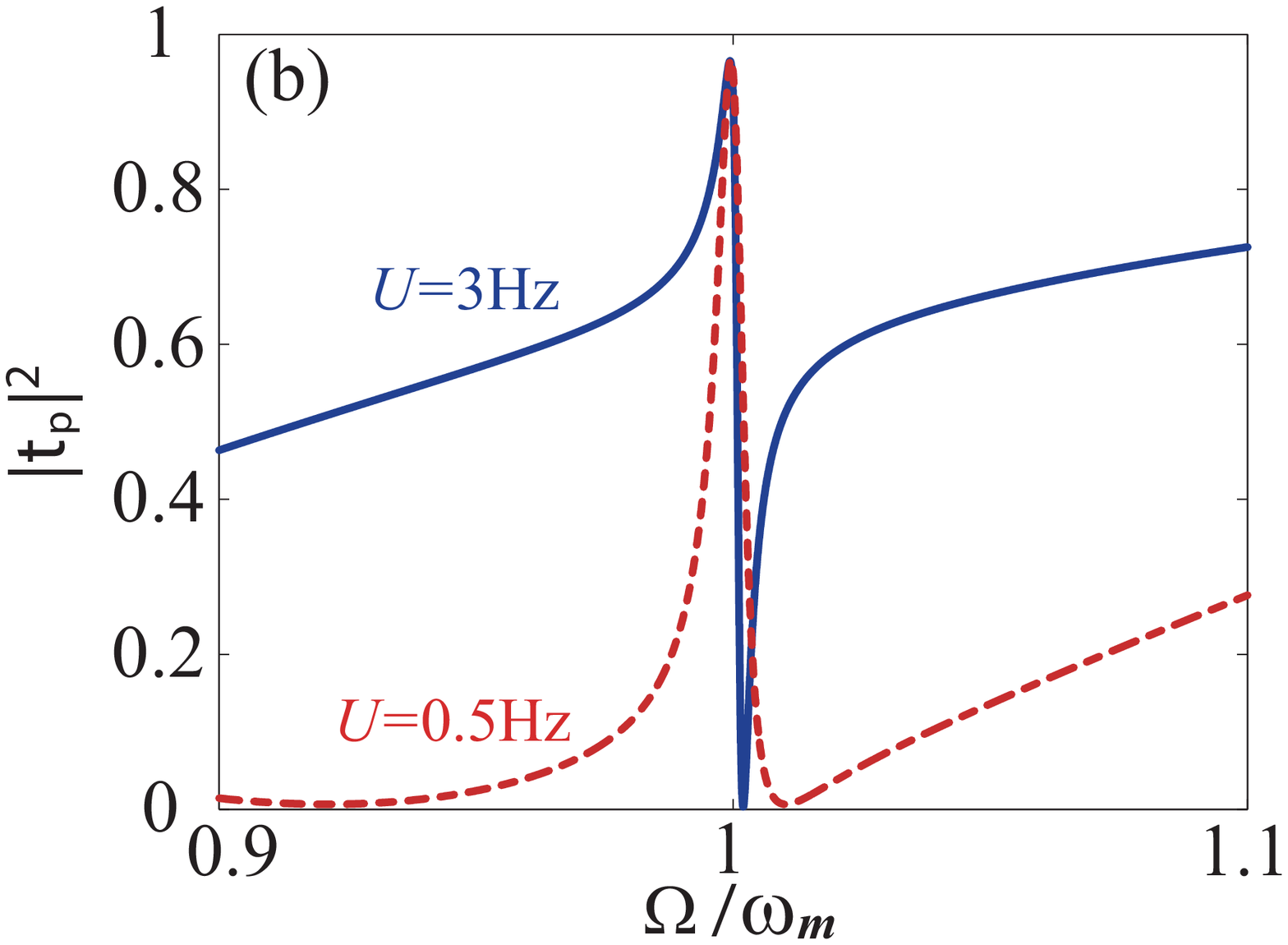}}
	\subfigure{
		\label{Fig2.sub.3}
		\includegraphics[width=0.32\textwidth]{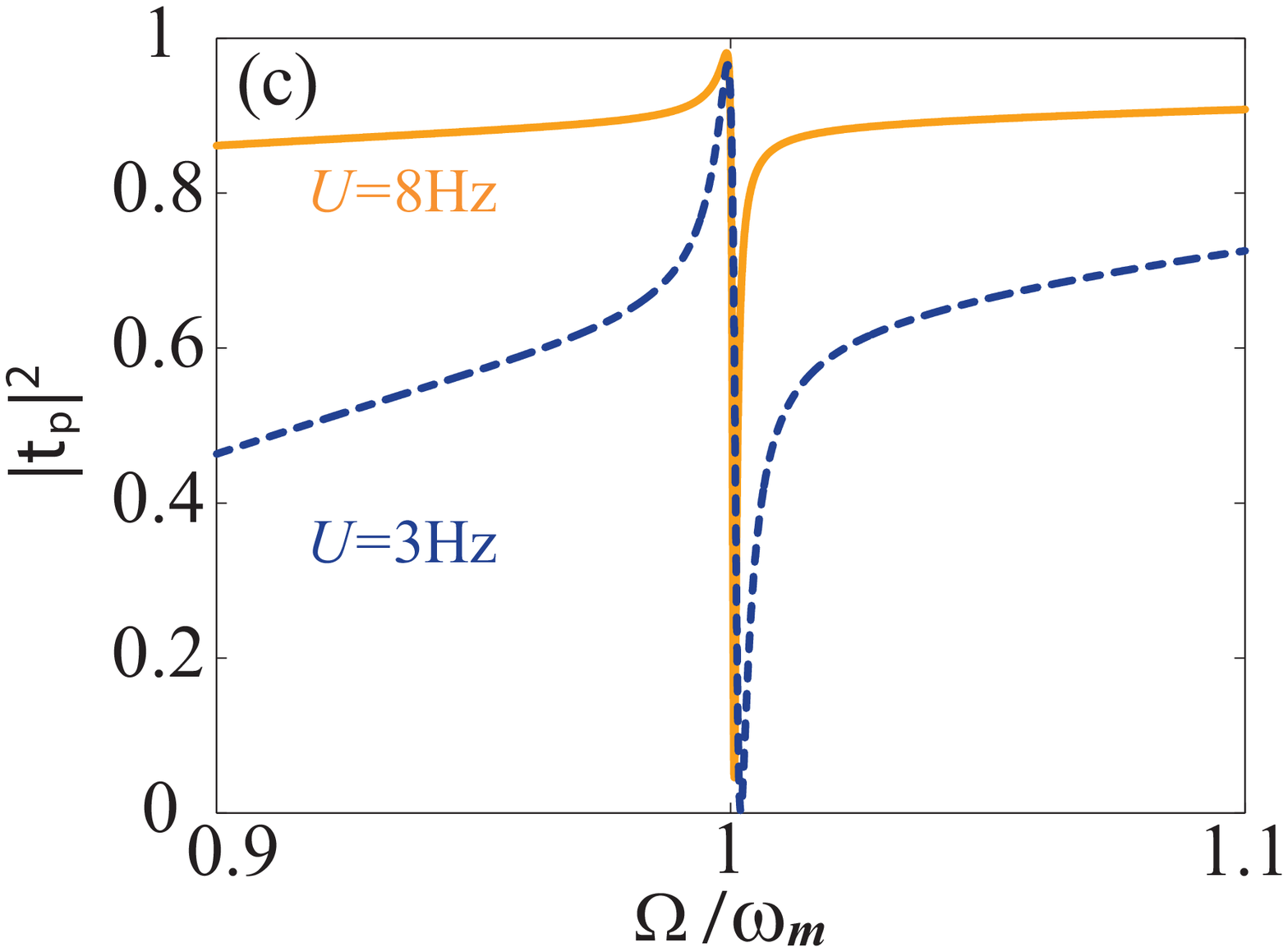}}
	\subfigure{
		\label{Fig2.sub.4}
		\includegraphics[width=0.32\textwidth]{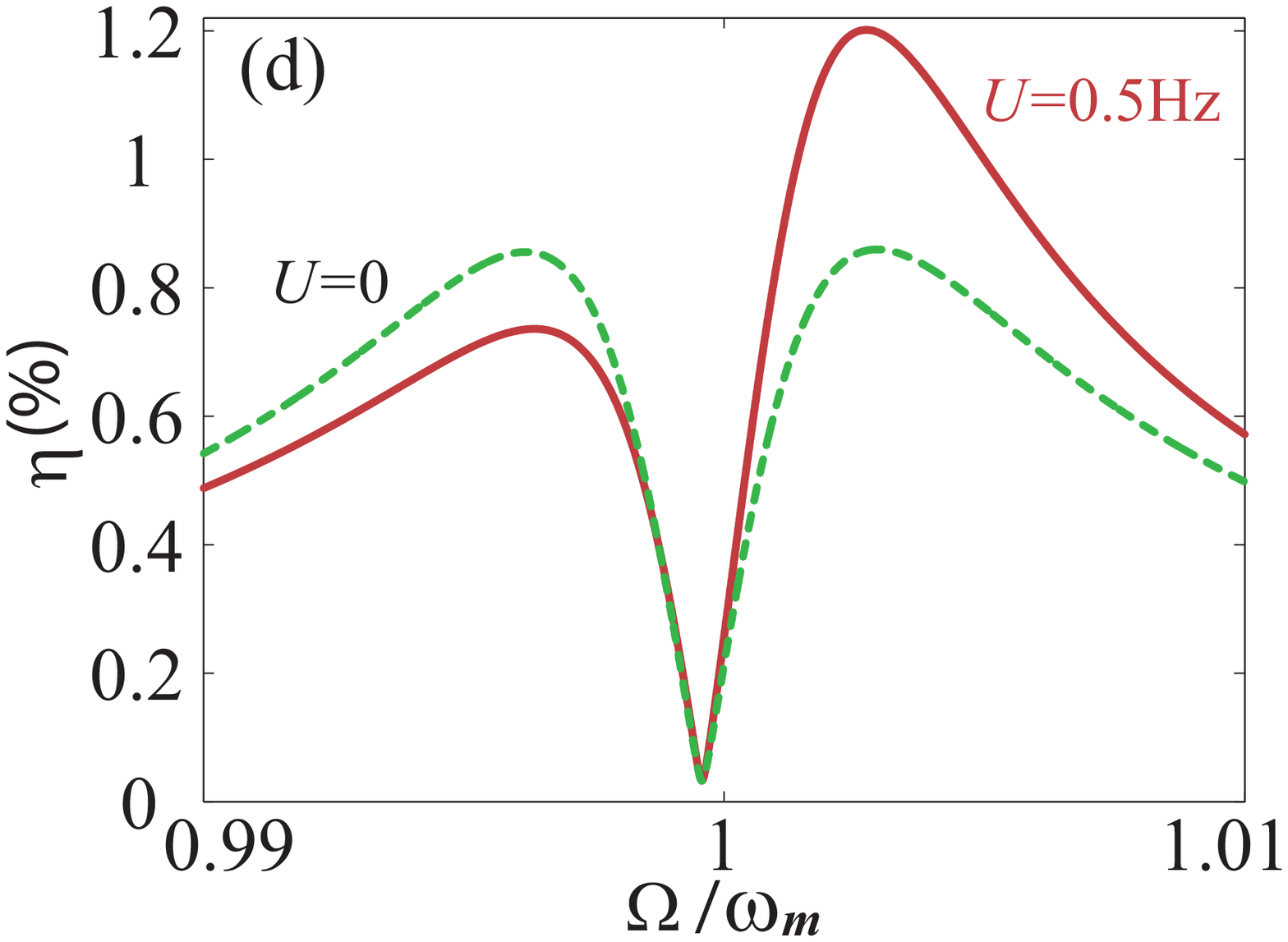}}
	\subfigure{
		\label{Fig2.sub.5}
		\includegraphics[width=0.32\textwidth]{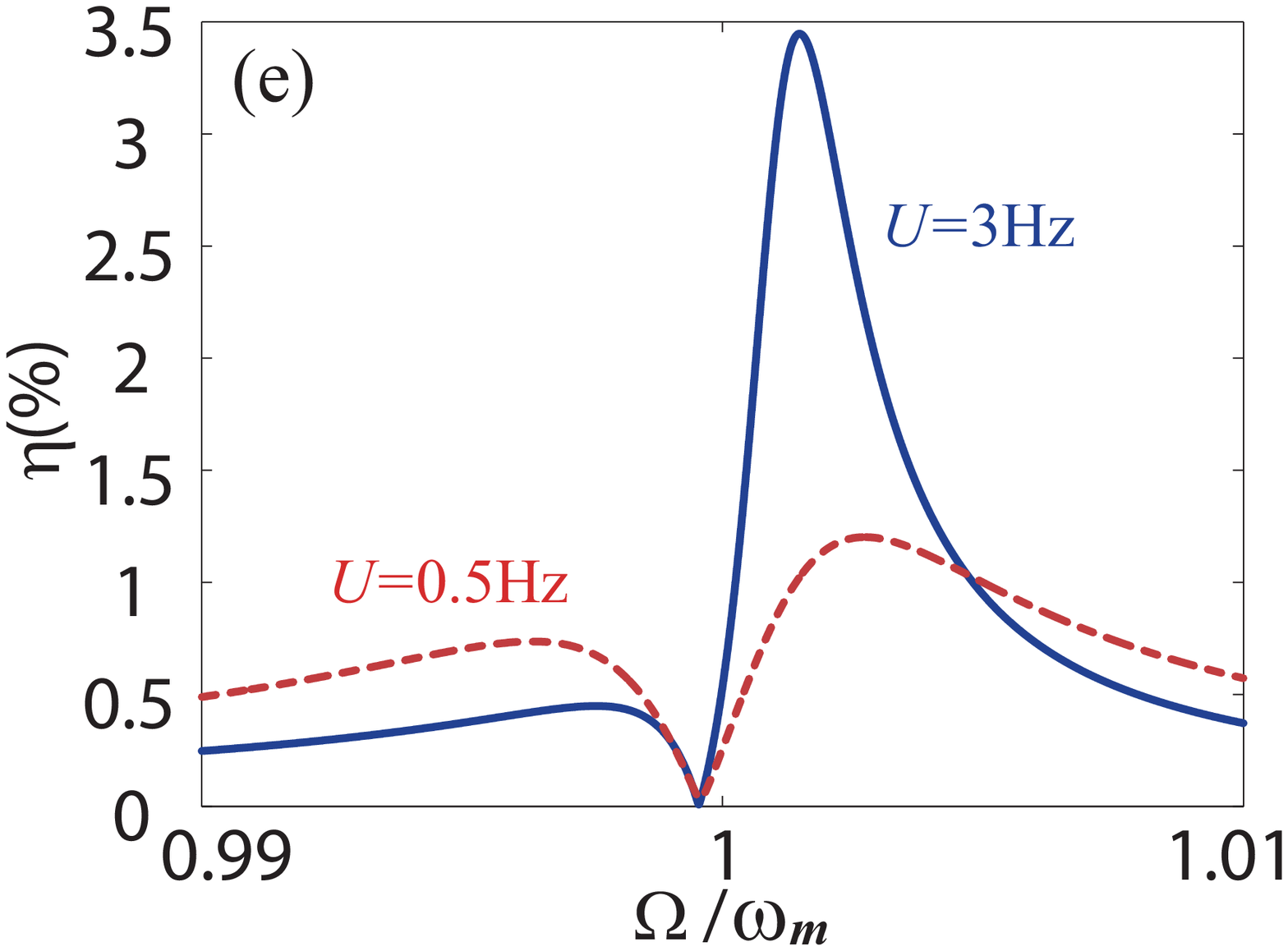}}
	\subfigure{
		\label{Fig2.sub.6}
		\includegraphics[width=0.32\textwidth]{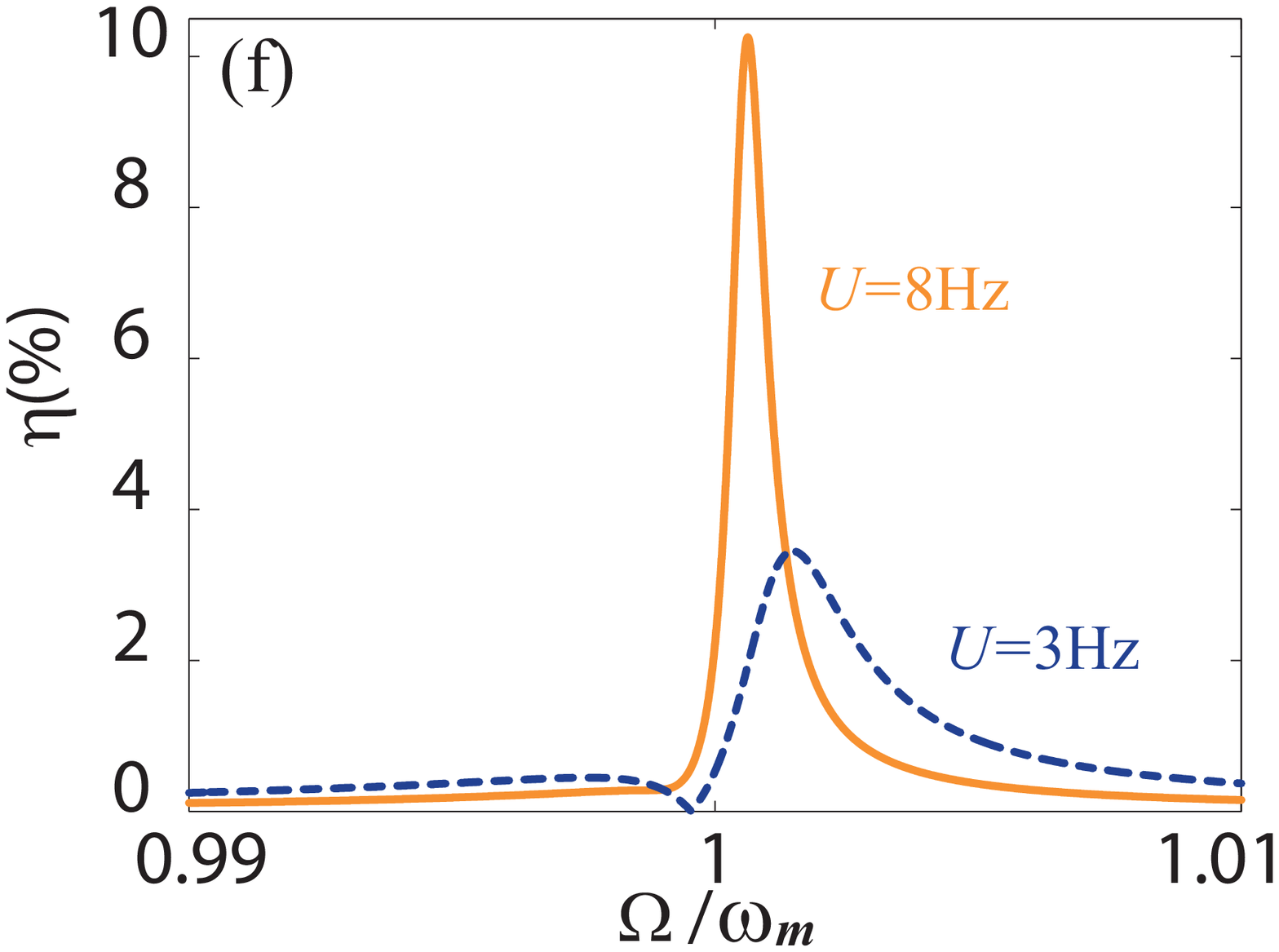}}	
	\caption{(Color online) Calculation results of the transmission rate of OMIT $ |t_p|^2 $ (a)--(c) and the efficiency of second-order sideband $ \eta $ (d)--(f) as a function of $ \Omega $ and $ U $, with pump power $ P_{L}=10\,\textrm{mW} $. Here $ U=0 $ indicates a linear resonator.}
	\label{Fig.2}
\end{figure*}
Since the probe field is much weaker than the control field, i.e., $\varepsilon_{p}\ll\varepsilon_{l}$, we can use the perturbation method to deal with Eq.\,(\ref{eq6}). Then, every operator can be expanded as the sum of its steady-state mean value and a small fluctuation around it, i.e., \(a=\bar{a}+\delta a(t)\) and \(x=\bar{x}+\delta x(t)\). We note that the steady-state mean values are much larger than the fluctuation terms, i.e., $|\bar{a}|\gg|\delta a(t)|$ and $|\bar{x}|\gg|\delta x(t)|$. Then the EOM of the fluctuation terms can be reduced as:
\begin{align}
\nonumber
&\frac{d}{dt}\delta a=(i\Delta-\kappa)\delta a-ig(\bar{a}\delta x+\delta a\delta x)\\ \nonumber
&~~~~~~~~~~~~~+2iU\bar{a}^2\delta a^\ast+\varepsilon_pe^{-i\Omega t},\\ \nonumber
&\delta\ddot{x}+\gamma_m\delta\dot{x}+\omega_m^2\delta x\\
&~~~~~~~~~=-\frac{g}{m}(\bar{a}\delta a^\ast+\bar{a}^\ast\delta a+\delta a\delta a^\ast),\label{eq7}
\end{align}
where 
\begin{align}
\nonumber
\Delta=\Delta_c-2\Delta\omega-g\bar{x}.
\end{align}
In Eq.\,(\ref{eq7}), the second-order terms, such as $ \delta a\delta x $ and $ \delta a\delta a^\ast$, are preserved to generate the required second-order sideband~\cite{Xiong860}. 

To calculate the amplitudes of the first-order and second-order sidebands, we assume that the fluctuation terms $ \delta a $ and $ \delta b $ have the following forms~\cite{Weis330}:
\begin{align}
\nonumber
&\delta a=A_{1}^{-}e^{-i\Omega t}+A_{1}^{+}e^{i\Omega t}+A_{2}^{-}e^{-2i\Omega t}+A_{2}^{+}e^{2i\Omega t}+\cdots,\\
&\delta x =X_{1}e^{-i\Omega t}+X_{1}^{\ast}e^{i\Omega t}+X_{2}e^{-2i\Omega t}+X_{2}^{\ast}e^{2i\Omega t}+\cdots.\label{eq8}
\end{align}
Here we only focus on the fundamental OMIT and its second-order sideband process, and thus the higher-order sidebands in Eq.\,(\ref{eq8}) are ignored. Substituting Eq.\,(\ref{eq8}) into Eq.\,(\ref{eq7}), we obtain six algebra equations which can be divided into two groups. The first group describes the linear response of the probe field,
\begin{align}
\nonumber
&\alpha_{-}A_{1}^{-}-2iU\bar{a}^{2}A_{1}^{+\ast}+ig(\bar{a}X_{1}+A_{1}^{+}X_{2}+X_{1}^{\ast}A_{2}^{-})=\varepsilon_{p},\\ \nonumber
&\alpha_{+}^{\ast}A_{1}^{+}-2iU\bar{a}^{2}A_{1}^{-\ast}+ig(\bar{a}X_{1}^{\ast}+A_{1}^{-}X_{2}^{\ast}+X_{1}^{\ast}A_{2}^{+})=0,\\ 
&\alpha_{0}X_{1}+g(\bar{a}A_{1}^{+\ast}+\bar{a}^{\ast}A_{1}^{-}+A_{1}^{+}A_{2}^{+\ast}+A_{1}^{-\ast}A_{2}^{-})=0,\label{eq9}
\end{align}
while the second group corresponds to the second-order sideband,
\begin{align}
\nonumber
&\lambda_{-}A_{2}^{-}-2iU\bar{a}^{2}A_{2}^{+\ast}+ig(\bar{a}X_{2}+A_{1}^{-}X_{1})=0,\\ \nonumber
&\lambda_{+}^{\ast}A_{2}^{+}-2iU\bar{a}^{2}A_{2}^{-\ast}+ig(\bar{a}X_{2}^{\ast}+A_{1}^{+}X_{1}^{\ast})=0,\\ 
&\lambda_{0}X_{2}+g(\bar{a}A_{2}^{+\ast}+\bar{a}^{\ast}A_{2}^{-}+A_{1}^{-}A_{1}^{+\ast})=0,\label{eq10}
\end{align}
where
\begin{align}
\nonumber
&\alpha_{\pm}=-i\Omega\pm i\Delta+\kappa,~~\alpha_{0}=m(\omega_{m}^{2}-i\gamma_{m}\Omega-\Omega^2),\\ \nonumber
&\lambda_{\pm}=-2i\Omega\pm i\Delta+\kappa,~\lambda_{0}=m(\omega_{m}^{2}-2i\gamma_{m}\Omega-4\Omega^2),\\
&\beta=-\dfrac{\big(2U\alpha_{0}+g^2\big)}{ig\big(\alpha_{+}+i\Delta\omega\big)}|\bar{a}|^2.\label{eq11}
\end{align}

Since the sideband amplitude is much smaller than that of the probe field, we can simplify Eq.\,(\ref{eq9}) by neglecting the higher-order nonlinear terms. Then, we can easily obtain the linear and nonlinear responses of the system as follows:

\begin{figure*}[ht]
	\centering
	\subfigure{
		\label{Fig3.sub.1}
		\includegraphics[width=0.32\textwidth]{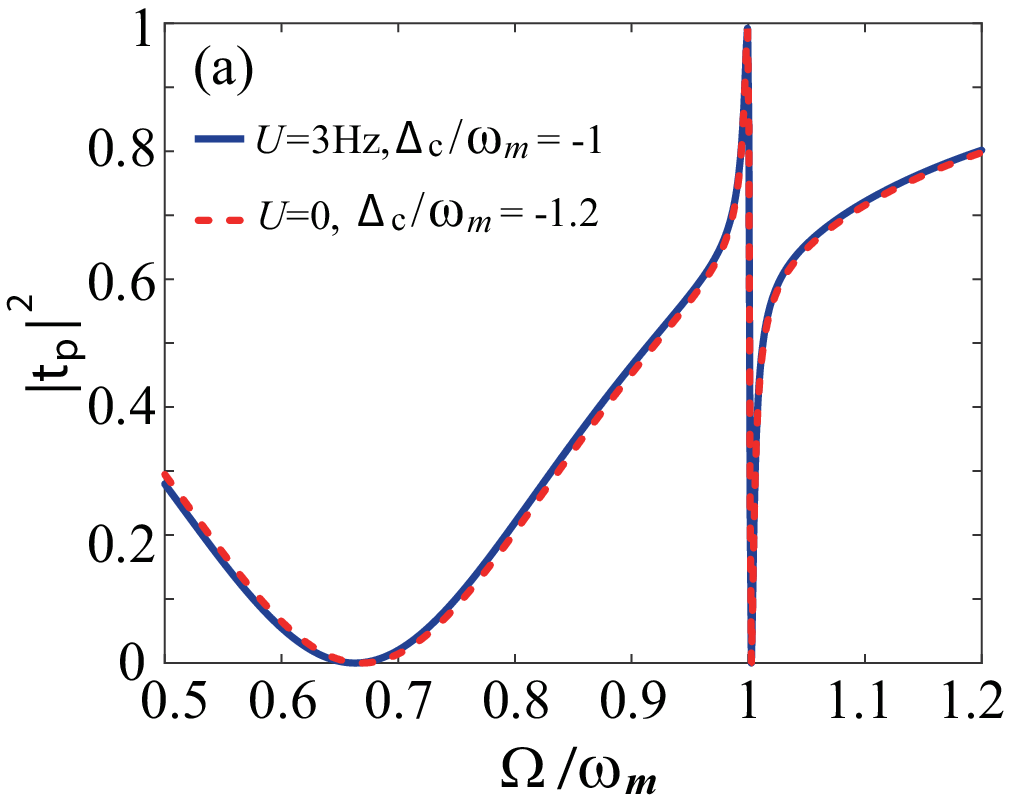}}
	\subfigure{
		\label{Fig3.sub.2}
		\includegraphics[width=0.32\textwidth]{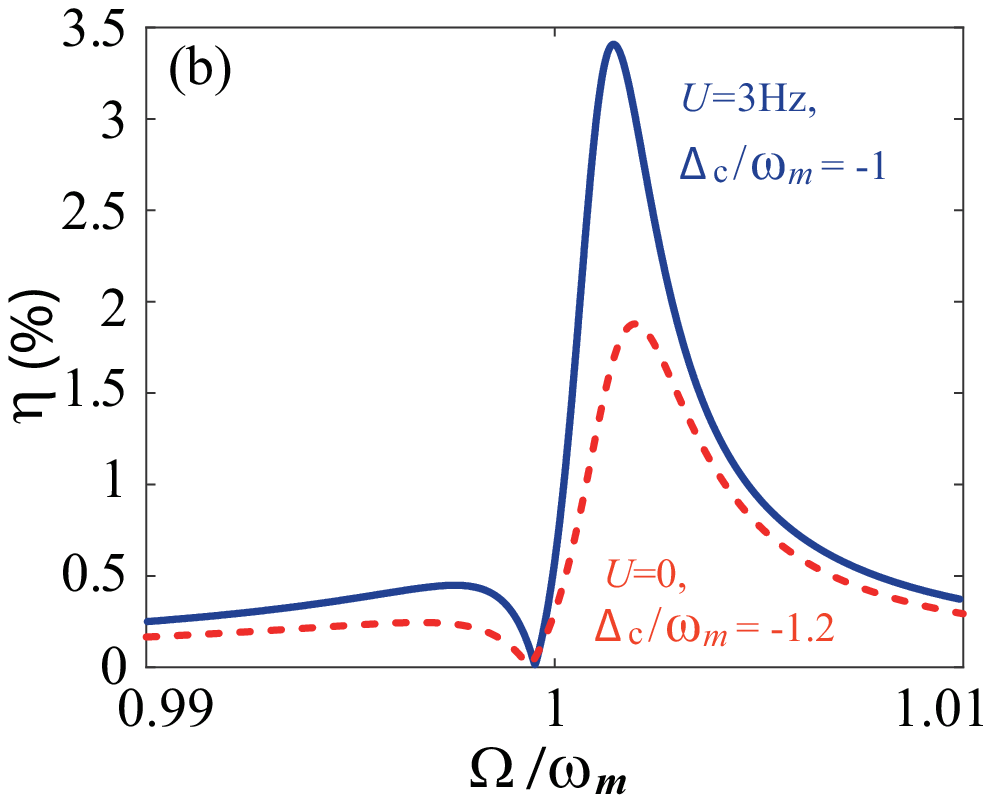}}
	\subfigure{
		\label{Fig3.sub.3}
		\includegraphics[width=0.32\textwidth]{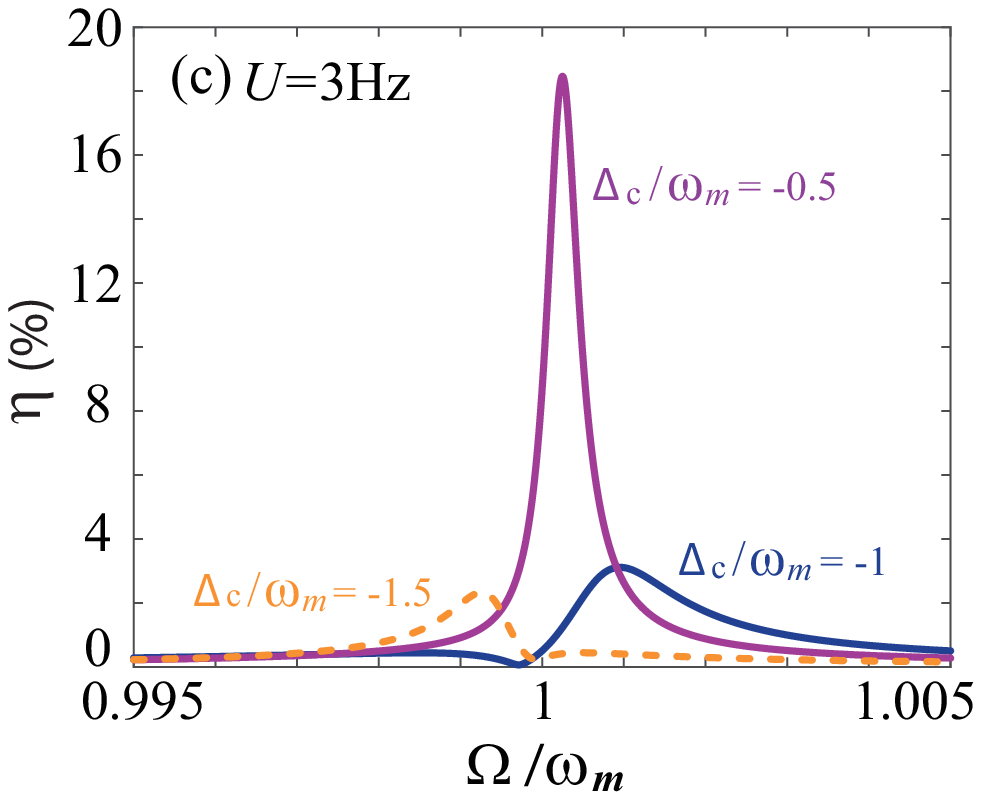}}	
	\caption{(Color online) Calculation results of the transmission rate of OMIT $|t_p|^2$ (a) and the efficiency of second-order sideband $ \eta $ (b)-(c) as a function of $ \Omega $, with pump power $ P_{L}=10\,\textrm{mW} $.}
	\label{Fig.3}
\end{figure*}

\begin{align}
\nonumber
&A_{1}^{-}=\dfrac{\alpha_{0}\alpha_{+}+ig^2|\bar{a}|^2}{\alpha_{0}\alpha_{+}\alpha_{-}-(\Delta\omega)^{2}\alpha_{0}+2g^2|\bar{a}|^2\big(\Delta+\Delta\omega\big)}\varepsilon_p,\\ \nonumber
\\ \nonumber
&X_{1}=\dfrac{-\big(\alpha_{+}+i\Delta\omega\big)g\bar{a}^\ast}{\alpha_{0}\alpha_{+}\alpha_{-}-(\Delta\omega)^{2}\alpha_{0}+2g^2|\bar{a}|^2\big(\Delta+\Delta\omega\big)}\varepsilon_p,
\end{align}
\begin{align}
\nonumber
A_{2}^{-}=&\dfrac{g\big(\lambda_{+}+i\Delta\omega\big)\beta-\big(\lambda_{0}\lambda_{-}+ig^2|\bar{a}|^2\big)}{\lambda_{0}\lambda_{+}\lambda_{-}-(\Delta\omega)^{2}\lambda_{0}+2g^2|\bar{a}|^2\big(\Delta+\Delta\omega\big)}A_{1}^{-}X_{1}\\
&+\dfrac{-\big(2U\lambda_{0}+g^2\big)\beta g\bar{a}^\ast }{\lambda_{0}\lambda_{+}\lambda_{-}-(\Delta\omega)^{2}\lambda_{0}+2g^2|\bar{a}|^2\big(\Delta+\Delta\omega\big)}X_{1}^2,\label{eq12}
\end{align}
where $A_{1}^{-}$ and $A_{2}^{-}$ are the coefficients of the first-order and second-order upper sidebands, respectively. It is obvious that $A_{2}^{-}$, being proportional to $|\varepsilon_{p}|^2$, is much smaller than $A_{1}^{-}$ for a weak probe laser. We also note that $A_{2}^{-}$ consists of two parts, i.e., the second-order sideband terms, being proportional to $(X_{1}^{-})^2$, and the upconverted first-order sideband terms. 

By using the standard input-output relations, i.e., 
\begin{align}
\nonumber
a_{out}=a_{in}-\sqrt{\kappa}a(t),
\end{align} 
we obtain the output fields of this OM system as follows:
\begin{align}
\nonumber
a_{out}&=s_{0}e^{-i\omega_lt}+s_{1}e^{-i\omega_pt}-\sqrt{\kappa}A_{2}^{-}e^{-i(2\omega_p-\omega_l)t}\\&~~~~-\sqrt{\kappa}A_{1}^{+}e^{-i(2\omega_l-\omega_p)t}-\sqrt{\kappa}A_{2}^{+}e^{-i(3\omega_l-2\omega_p)t},\label{eq13}
\end{align}
with
\begin{align}
s_{0}=\varepsilon_l/\sqrt{\kappa}-\sqrt{\kappa}\bar{a},~~~~s_{1}=\varepsilon_p/\sqrt{\kappa}-\sqrt{\kappa}A_{1}^{-}.\label{eq14}
\end{align}
The term $ s_{0}e^{-i\omega_lt} $ denotes the output with pump frequency $ \omega_{l} $, while the terms $-\sqrt{\kappa}A_{1}^{+}e^{-i(2\omega_l-\omega_p)t}$ and $s_{1}e^{-i\omega_pt}$ describe the Stokes and anti-Stokes fields, respectively. Moreover, the terms $-\sqrt{\kappa}A_{2}^{+}e^{-i(3\omega_l-2\omega_p)t}$ and $-\sqrt{\kappa}A_{2}^{-}e^{-i(2\omega_p-\omega_l)t}$, describing the output with frequencies $ \omega_{l}\pm2\Omega $, are related to the second-order upper and lower sidebands, respectively~\cite{Xiong860}. Here, we only consider upper sidebands, and one can use the same method to discuss the lower ones. Then, the transmission rate of the probe field or the efficiency of the second-order upper sideband can be defined as:

\begin{align}
|t_p|^2=|1-\frac{\kappa}{\varepsilon_p} A_{1}^{-}|^2,~~~~~~\eta=|-\frac{\kappa}{\varepsilon_p}A_{2}^{-}|.\label{eq15}
\end{align}

The associated group delay of the probe light or the second-order upper sideband is given by~\cite{Safavi472}:
\begin{align}
\tau_1=\dfrac{d \arg(t_p)}{d \Omega}|_{\Omega=\omega_m},~~~~~~\tau_2=\dfrac{d \arg(\eta)}{2d \Omega}|_{\Omega=\omega_m}.\label{eq16}
\end{align}

In our numerical simulations, to demonstrate that the observation of the second-order sideband in a Kerr resonator is within current experimental reach, we calculate Eqs.\,(\ref{eq15}) and (\ref{eq16}) with parameters from Ref.~\cite{Shen411}: $\lambda=780\,\textrm{nm}$, $m=50\,\textrm{ng}$, (the resonator radius) $R=19\,\mu m$, $n_{0}=1.47$, $n_{2}=3.2\times10^{-16}\,\textrm{cm}^{2}/\textrm{W}$, $ Q=\omega_{c}/\kappa=1.7\times10^{7} $,  $\omega_{m}/2\pi=83.7\,\textrm{MHz}$, $\gamma_{m}/2\pi=20\,\textrm{kHz}$, $\kappa/2\pi=22\,\textrm{MHz}$, $\varepsilon_{p}/\varepsilon_{l}=0.05$, $ g=-\omega_{c}/R $, and $\Delta_{c}=-\omega_{m}$, respectively.

In Fig.\,\ref{Fig.2}, the transmission rate $ |t_p|^{2} $ and the efficiency $ \eta $ are shown as a function of the detuning $ \Omega $. In linear OM system, the OMIT transmission window is symmetric and located around $ \Omega\,/\,\omega_{m}=1 $~\cite{Agarwal810,Weis330,Safavi472}. In contrast, the second-order sideband demonstrates a local minimum between the two sideband peaks around the resonance $ \Omega\,/\,\omega_{m}=1 $ [see the dashed line in Fig.\,\ref{Fig2.sub.4}]. The efficiency of second-order sideband is usually very small in the OM system, e.g., $ 1\%-2\% $, depending on the pump power and OM coupling rate. This is because the generation of the second-order sideband is mainly from the upconverted first-order sideband. Under the condition $ \Omega/\omega_{m}=1 $ and $ \Delta/\omega_{m}=1 $, the anti-Stokes field is resonantly enhanced, thus leading to the suppression of the second-order sideband. As shown in Figs.\,\ref{Fig2.sub.1}--\ref{Fig2.sub.3}, the OMIT spectrum becomes asymmetric in the presence of the Kerr effect. The transparency window of OMIT is not at the center and its linewidth broadens with the increasing $ U $. The reason is that the linewidth of the OMIT window is related to the intracavity photon number~\cite{Weis330}, i.e., 
\begin{equation}
\Gamma=\gamma_{m}+\dfrac{(gx_{zpf})^{2}}{\kappa}|\bar{a}|^{2},
\end{equation}
where $ x_{zpf}\equiv\sqrt{\hbar/2m\omega_{m}} $ is the zero-point fluctuations of the mechanical mode. The Kerr effect, significantly increasing the intracavity photon number, induces a red shift of the cavity mode by $ \Delta\omega $. Moreover, the modified value of $ |\bar{a}|^{2} $ broadens the linewidth and leads to a Fano-like OMIT spectrum. We show the efficiency $ \eta $ under different Kerr strengths in Figs.\,\ref{Fig2.sub.4}--\ref{Fig2.sub.6}. Compared to that in a linear OM system, the second-order sideband is significantly enhanced in a Kerr resonator. For example, the efficiency $ \eta $ is about $ 10\% $ for $ U=8\,\textrm{Hz} $, i.e., 3 times that for $ U=3\,\textrm{Hz} $. 
\begin{figure}[bt]
	\centering
	\includegraphics[width=0.4\textwidth]{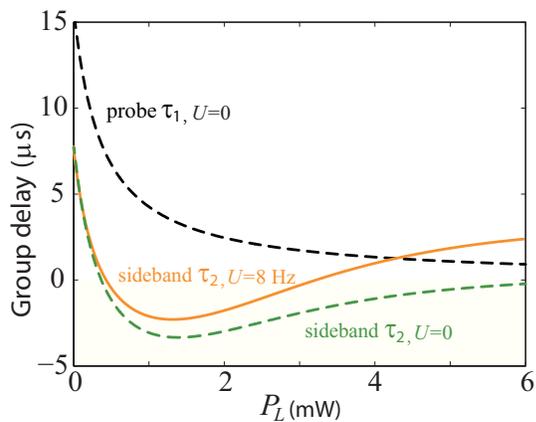}	
	\caption{(Color online) Optical group delay as a function of the pump power $ P_{L} $ at the resonance $ \Omega/\omega_{m}=1 $.}\label{Fig.4}
\end{figure}

Figure\,\ref{Fig.3} shows $ |t_p|^{2} $ and $ \eta $ as a function of $ \Omega $ under the condition that pump-cavity detuning $ \Delta_{c} $ and mechanical frequency $ \omega_{m} $ are off resonance. The results of $ \eta $ by either adjusting pump frequency or steering the Kerr strength are shown in Figs.\,\ref{Fig3.sub.1} and \ref{Fig3.sub.2}. The transmission windows of OMIT are almost the same, while the efficiency of second-order sideband under these two conditions can be quite different. This can be explained as follows: On one hand, steering the Kerr strength or adjusting pump frequency changes the value of $ \Delta_{c} $, hence the anti-Stokes field is no longer resonantly enhanced, and the upconverted process of the first-order sideband is strengthened. On the other hand, besides OM interaction, the second-order sideband can also emerge from the Kerr nonlinearity. Consequently, the enhancement of $ \eta $ by the Kerr effect is higher than that by adjusting pump frequency. As shown in Fig.\,\ref{Fig3.sub.3}, $ \eta $ can be further enhanced by increasing pump frequency in a Kerr resonator. For example, $ \eta $ can reach about $ 20\% $ for $ U=3\,\textrm{Hz} $ and $ \Delta_c/\omega_{m}=-0.5 $. However, the efficiency decreases when decreasing the pump frequency [see the dashed line in Fig.\,\ref{Fig3.sub.3}]. This is because the Kerr-induced frequency shift can be either compensated or amplified by adjusting the pump frequency.   

Finally, associated with the OMIT, the optical group delay can emerge in such a system. This is similar to that of EIT, experiencing a dramatic reduction in its group velocity resulting from the rapid variation of the refractive index. To see this, we plot the corresponding group delay of the transmitted light in Fig.\,\ref{Fig.4}. In a conventional OM system, the delay of the transmitted light is only related to the pump power~\cite{Safavi472}. With the increasing power, both the probe light and the second-order sideband tend to be advanced. The second-order sideband even switches to fast light. However, the delay time of the second-order sideband increases at high pump power in the presence of the Kerr effect, which is useful for the storage. We note that in the presence of the Kerr effect, the refractive index of the material depends on the intensity of the cavity field $ I $, i.e., $ n(I)=n_{0}+n_{2}I $. As mentioned before, the intracavity photon number is greatly influenced by the Kerr effect, therefore, the group delay could be tuned by adjusting the nonlinear Kerr coefficient at high pump power.

In summary, we have theoretically studied the second-order OMIT sidebands in a Kerr resonator. In such a resonator, due to its high Q factor and small mode volume, the nonlinear Kerr effect can be considerably enhanced as experimentally confirmed~\cite{Cao118,Mitchell425,Shen411}. A redshift of the cavity mode and thus an asymmetric OMIT spectrum were already observed in a very recent experiment~\cite{Shen411}. Here we find that the OMIT second-order sidebands also can be largely enhanced, since the upconverted process of the first-order sideband and the Kerr-induced second-order sideband can be both enhanced in this system. In addition, the group delay of the second-order sidebands in a Kerr resonator is revealed to be capable of switching from fast light to slow light by tuning the external light. These results are helpful to better understand the propagation of light in nonlinear OM devices, indicating that a wide range of OM effects could be further steered with various optical nonlinearities. Other future developments along this line may include OMIT with nonlinear parity-time resonators~\cite{Jing596}, nonreciprocal devices with Kerr resonators~\cite{Cao118,Shen106}, or topological effects with an array of Kerr resonators.

We are grateful to Chun-Hua Dong at the University of Science and Technology of China for his insightful discussions. This work is supported by the National Natural Science Foundation of China of China under Grants No. 11474087 and No. 11774086, and the HuNU Program for Talented Youth.

%\begin{align}
%\nonumber
%&\alpha_{-}A_{1}^{-}-2iU\bar{a}^{2}A_{1}^{+\ast}+ig(\bar{a}X_{1}+A_{1}^{+}X_{2}+X_{1}^{\ast}A_{2}^{-})=\varepsilon_{p},\\ \nonumber
%&\alpha_{+}^{\ast}A_{1}^{+}-2iU\bar{a}^{2}A_{1}^{-\ast}+ig(\bar{a}X_{1}^{\ast}+A_{1}^{-}X_{2}^{\ast}+X_{1}^{\ast}A_{2}^{+})=0,\\ \nonumber
%&\alpha_{0}X_{1}+g(\bar{a}A_{1}^{+\ast}+\bar{a}^{\ast}A_{1}^{-}+A_{1}^{+}A_{2}^{+\ast}+A_{1}^{-\ast}A_{2}^{-})=0,\\ \nonumber
%&\lambda_{-}A_{2}^{-}-2iU\bar{a}^{2}A_{2}^{+\ast}+ig(\bar{a}X_{2}+A_{1}^{-}X_{1})=0,\\ \nonumber
%&\lambda_{+}^{\ast}A_{2}^{+}-2iU\bar{a}^{2}A_{2}^{-\ast}+ig(\bar{a}X_{2}^{\ast}+A_{1}^{+}X_{1}^{\ast})=0,\\ 
%&\lambda_{0}X_{2}+g(\bar{a}A_{2}^{+\ast}+\bar{a}^{\ast}A_{2}^{-}+A_{1}^{-}A_{1}^{+\ast})=0. \label{eq9}
%\end{align}

\end{document}